# A Trustability Metric for Code Search based on Developer Karma

Florian S. Gysin
Software Composition Group
University of Bern, Switzerland
flo.g@students.unibe.ch

Adrian Kuhn
Software Composition Group
University of Bern, Switzerland
http://scg.unibe.ch/akuhn

## ABSTRACT

The promise of search-driven development is that developers will save time and resources by reusing external code in their local projects. To efficiently integrate this code, users must be able to trust it, thus *trustability* of code search results is just as important as their relevance. In this paper, we introduce a *trustability metric* to help users assess the quality of code search results and therefore ease the cost-benefit analysis they undertake trying to find suitable integration candidates. The proposed trustability metric incorporates both user votes and cross-project activity of developers to calculate a *"karma"* value for each developer. Through the karma value of all its developers a project is ranked on a trustability scale. We present *JBender*, a proof-of-concept code search engine which implements our trustability metric and we discuss preliminary results from an evaluation of the prototype.

## 1. INTRODUCTION

Code search engines help developers to find and reuse software. However, to support search-driven development it is not sufficient to implement a mere full text search over a base of source code, human factors have to be taken into account as well. At last year's SUITE workshop [9], *suitability* and *trustability* have been major issues in search-driven development, besides—of course—relevance of search results.

In this paper we focus on the *trustability* of search results. Relevance of code search results is of course paramount, but trustability in the results is just as important. Before integrating a search result the developer has to assess its trustability to take a go-or-no-go decision. A well-designed search interface allows its users to take this decision on the spot. Gallardo-Valencia et al. found that developers often look into human rather than technical factors to assess the trustability of search results [4]. For example developers will prefer results from well-known open source projects over results from less popular projects.

In this paper we present a trustability metric for search results. The trustability metric is based on human factors. We use data collected from Web 2.0 platforms to assess the trustability of both projects and developers. Our trustability metric is based on collaborative filtering of user votes and cross-project activity of developers. For example, if a little-known project is written by developers who also contributed to a popular open source project, the little-known project is considered to be as trustable as the popular project.

As a feasibility study, we implemented the trustability metric in *JBender*, a proof-of-concept code search engine. The index of our *JBender* installation currently contains trustability assessments for over 3,700 projects, based on 193,000 user votes and the cross-project activity of over 56,000 developers. In this paper, preliminary results from an evaluation of the prototype are discussed.

The remainder of this paper is structured as follows. Section 2 discusses background and related work. Section 3 introduces our trustability metric, and Section 4 describes *JBender*, a proof-of-concept prototype. Section 5 discusses preliminary results from an ongoing evaluation of the prototype. Eventually, we conclude in Section 6 with remarks on future work.

## 2. BACKGROUND & RELATED WORK

Since the rise of internet-scale code search engines, searching for reusable source code has quickly become a fundamental activity for developers [1]. However, in order to establish search-driven software reuse as a best practice, the cost and time of deciding whether to integrate a search result must be minimized. The decision whether to reuse a search result or not should be quickly taken without the need for careful (and thus time-consuming) examination of the search results.

Trustability is a big issue for reusing code. When a developer reuses code from an external sources he has to trust the work of external developers that are unknown to him. This is not to be confused with trustworthy computing, where clients are concerned with security and reliability of a computation service.

For a result to actually be helpful and serve the purpose originally pursued with the search it is not enough to just match the entered keywords. User studies have shown that developers rely on both technical and human clues to assess the trustability of search results [4]. For example developers will prefer results from well-known open source projects over results rom less popular projects.

The issue of providing meta-information alongside search results and thereby increasing trustabilty has not been widely studied and we are trying to address this with our work.

In recent years special search engines for source code have appeared, namely GOOGLE CODE SEARCH [1], KRUGLE [2] and KODERS [3]. They all focus on full-text search over a huge code base, but lack detailed information about the project. Search results typically provide a path to the version control repository and little meta-information on the actual open source project; often, even such basic information as the name and homepage of the project are missing.

SOURCERER [4] by Bajracharya et al. [2] and MEROBASE [5] by Hummel et al. [6] are research projects with an internet-scale code search-engine. Both provide the developer with license information and project name. Merobase also provides a set of metrics such as cyclomatic- and Halstead complexity. An improved version of Sourcerer with trustability data is in development, though it has not yet been published[6].

In addition to the web user interface, both Sourcerer and Merobase are also accessible through Eclipse plug-ins that allow the developer to write unit tests. These are then used as a special form of query to search for matching classes/methods, *i.e.*, classes that pass the tests[6]. Using unit tests as form of formulating queries is a way of increase technical trustability: Unit-tested search results are of course more trustable, however at the cost of a more time consuming query formulation (*i.e.*, additionally writing the unit tests). The kind of results returned are also limited to clearly-defined and testable features. A combination of technical trustability factors (*e.g.*, unit tests) and human trustability factors might be promising future work.

We are not the first to use collaborative filtering in code search. Ichii et al. used collaborative filtering to recommend relevant components to users [7]. Their system uses browsing history to recommend components to the user. The aim was to help users make cost-benefit decisions about whether or not those components are worth integrating. Our contribution beyond the state-of-the-art is our focus on human factors and the role of cross-project contributors.

## 3. TRUSTABILITY METRIC

In this section, we propose a trustability metric for code search results that uses collaborative filtering of both user votes and cross-project activity of developers.

To assess the trustability of code search results we combine traditional full text search with meta-information from Web 2.0 platforms. Our trustability metric requires the following information:

- A matrix $M = (c_{d,p})$ with the number of contributions per contributor $d$ to a project $p$.

- A vector $V = (v_p)$ with user votes for software projects to signal the users' trust in projects. Gallardo-Valencia et al. refer to user votes as "fellow users" [4].

We use collaborative filtering of both user votes and cross-project activity of developers. For example, if a little-known project is written by developers who have also contributed to a popular open source project, the little-known project is considered to be as trustable as the popular project. Since both the number of contributions per contributor and the number of votes per project follow a power-law distribution, we use *log* weighting and *tf-idf* [7] weighting where applicable.

---

[1] http://www.google.com/codesearch
[2] http://www.krugle.org
[3] http://www.koders.com
[4] http://sourcerer.ics.uci.edu
[5] http://www.merobase.org
[6] Personal communication with Sushil Bajracharya.

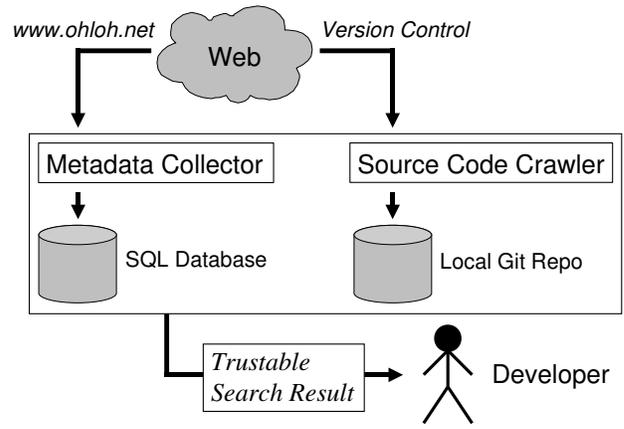

**Figure 1:** Architecture of the *JBender* prototype. *JBender* enhances search results from source code with a trustability estimate that is based on social data collect from the Ohloh Web 2.0 website.

First we define the *karma* of a contributor as

$$K_d = \sum_P w_{d,p} \log v_p \quad \text{where} \quad w_{d,p} = \frac{\log c_{d,p}}{\log \mathrm{df}(d)}$$

which is the sum of the votes of all projects, weighted by the number of contributions to these projects and divided by the inverse project frequency of the contributor (*i.e.*, the number of projects to which the contributor contributed at least one contribution).

Based on this, trustability of a project is defined as

$$T_p = \sum_D w_{d,p} K_d \quad \text{where} \quad w_{d,p} = \frac{\log c_{d,p}}{\sum_{d' \in D} \log c_{d',p}}$$

which represents the sum of the karma of all the projects contributors, weighted by the number of their contributions. Note that we divide project trustability by the total number of contributions, but not contributor karma. This is on purpose, contributors are more trustable the more they commit (based on the assumption that all accepted commits require approval of a trusted core developer, as is common in many open source projects) but projects are not per se more trustable the larger they are.

To summarize, we consider a project to be trustable if there are significant contributions by contributors who have also significantly contributed to projects (including the project in question) that have received a high number of user votes.

The proposed definition of trustability is dominated by cross-project contributors, *i.e.*, contributors who contributed many times to many projects with many votes. This is in accordance with empirical findings on open source that have shown how cross-project developers are a good indicator of project success [8]. This behaviour is also known as "the rich get richer" in the theory of scale-free networks and is considered an inherent and thus common property of most social networks [3].

---

[7]"term frequency-inverse document frequency" http://en.wikipedia.org/wiki/Tf-idf

| Project Trustability | Info |
|---|---|
| Project: **JUnit** | **ArrayComparisonFailure** extends *AssertionError* in *org.junit.internal* |
| **Trustability: 26.87**<br>License: CPL<br>856 users<br>7 developers | ```<br>private final AssertionError fCause;<br><br>/**<br> * Construct a new ArrayComparisonFailure with an error text and the array's<br> * dimension that was not equal<br> * @param cause the exception that caused the array's content to fail the assertion test<br> * @param index the array position of the objects that are not equal.<br>``` |

**Figure 2:** Screenshot of a *JBender* search result with trustability estimate. On the right there is the actual search result, with full name and code snippet. On the left there is information about the originating project and the trust value calculated by the trustability metric.

## 4. THE JBENDER PROTOTYPE

We have developed a prototype, called *JBender*, which enriches code search results with trustability information. To add to the information content of search results we combine two main sources to form the *JBender* code search engine. On the one hand there is the actual code base of the search engine over which an index is created. On the other hand we have created a database of metadata for the projects in the code base.

Figure 1 illustrates the architecture of *JBender*. *JBender* creates a searchable index over the code base and provides a code search over it. Its novelty however lies in the underlying metadata which is linked to the projects in the searchable code base - upon finding results from the latter *JBender* can supply the meta information stored for the result's originating project.

### 4.1 JBender's Metadatabase

Our source of meta data is the OHLOH[8] project. Ohloh is a social networking platform for open source software projects where projects (or rather their developers) can specify additional information. However Ohloh does not allow users to actually search through or interact with the source code: Ohloh is not a code search engine. Ohloh provides user contributed information on both open source projects and their developers, composing valuable information for search users. Users can vote for both projects and developers whether and how much they like them by rating projects and giving kudos to certain developers. Furthermore kudos are (automatically) given to developers who have worked for successful projects, i.e. projects with large user bases.

For the *JBender* prototype we collected the trustability meta-information from Ohloh, which is a social web platform for open source projects that provides user contributed information on both open source projects and their developers.

Metadata stored in the database includes (among others): Description of original project, project homepage, rating of the project, list of current repositories (type, url, last time of update, ...), licenses of files in the project (exact type of license, number of files), employed programming languages (percentage of total, lines of code, comment ratio, ...), the project's users and developers who worked on the project (kudos, experience, commits per project, ...).

### 4.2 JBender's Codebase

In addition to the collected metadata, *JBender* also follows the links to the version control repositories that are listed on Ohloh, creates local copies of these repositories and parses the code in Java projects to build an search index over them. *JBender* then provides a basic structured code search over various parts of the indexed source code. Examples are method/class names and their bodies, comments, visibility, dependencies and implemented interfaces.

### 4.3 Trustability enhanced results

The following data from Ohloh was directly used for the trustability metric: As contributors we used the developers of the projects and as the number of contributions we used the number of commits. As user votes we used the number of developers who "stacked" a project, which is Ohloh's terminology for claiming to be an active user of a project.[9] Thus in our case, both users and contributors are open source developers. To be a user the developers must be registered on Ohloh. This is not necessary for being a contributor, since that information is taken from version control systems.

As explained in Section 3 this trustability metric takes into account several of the collected meta parameters and calculates a trust metric for each result according to which the results can be sorted.

Figure 2 shows a screenshot of a single search result from *JBender*. On the right there is the actual search result, with full name and code snippet. On the left there is information about the originating project and the trust value calculated by the trustability metric. Currently the raw trust measurement is displayed as a floating point number to the user. We might change that to a ranked assessment that maps the trustability to a scale from 1 to 10 to improve usability.

The layout of our search result is deliberately kept very simple and lucid in order to be efficiently usable. It has been shown that efficient search requires compact and well-arranged interfaces, which do not burden the user with too much information or a complex information seeking process [5].

---

[8] http://www.ohloh.net

[9] That is, we interpret "votes" as a user expressing his trust in a project by using it.

| Top projects (by votes) | Top Developer (by karma) | Top projects (by trustability) |
|---|---|---|
| "firefox", $v_p = 7207$ | "darins", $K_d = 71.97$ | "grepWin", $T_p = 51.60$, $v_p = 32$ |
| "subversion", $v_p = 5687$ | "amodra", $K_d = 70.11$ | "GNU Diff Utilities", $T_p = 51.18$, $v_p = 645$ |
| "apache", $v_p = 5107$ | "darin", $K_d = 69.09$ | "Eclipse Ant Plugin", $T_p = 49.76$, $v_p = 136$ |
| "mysql", $v_p = 4834$ | "nickc", $K_d = 67.14$ | "Eclipse Java Development Tools", $T_p = 48.36$, $v_p = 647$ |
| "php", $v_p = 4081$ | "Dani Megert", $K_d = 66.51$ | "Crimson", $T_p = 42.41$, $v_p = 2$ |
| "openoffice", $v_p = 3118$ | "mlaurent", $K_d = 66.14$ | "GNU binutils", $T_p = 42.18$, $v_p = 525$ |
| "firebug", $v_p = 3109$ | "Paul Eggert", $K_d = 65.89$ | "syrep", $T_p = 42.12$, $v_p = 2$ |
| "gcc", $v_p = 2586$ | "kazu", $K_d = 65.78$ | "GNU M4", $T_p = 41.85$, $v_p = 54$ |
| "putty", $v_p = 2519$ | "rth", $K_d = 65.25$ | "gzip", $T_p = 41.61$, $v_p = 261$ |
| "phpmyadmin", $v_p = 2412$ | "hjl", $K_d = 65.04$ | "Forgotten Edge OpenZIS", $T_p = 40.86$, $v_p = 1$ |

**Figure 3:** Top ten results for A) project ranking by Ohloh, B) karma of developers, C) project ranking by trustabilty.

## 5. DISCUSSION

*Some preliminary results.*

Figure 3 illustrates the top-10 results for a) project ranking through votes by Ohloh, b) karma of developers, c) project ranking by our trustability metric. Notice how the project ranking changed through consideration of cross-project developer activity: grepWin for example has only 32 users on Ohloh but is ranked by us with top trustability because its developers are very active and have a high karma value.

*Evidence of power law distribution.*

We found that our input data (*i.e.*, the user-generated data that we crawled from Ohloh) follows a power law distribution: the number of votes per project ($r = 0.95157$), the number of commits per developer per project ($r = 0.89207$), as well as number of projects per developer ($r = 0.85029$). Therefore we applied *log* and *tf-idf* weighting so that the trustability metric is not dominated by high values. At the moment project trustability ranges from zero to about 52, developer karma ranges from zero to about 72.

*A note on Ohloh's kudo-rank.*

The Ohloh website provides its own measurement of developer "karma", called *kudo-rank*. Kudo-ranks are based on a mix of user votes for projects and of user votes for developers, called *kudos*. User participation for kudos is very low and as a consequence a small clique of developers can vote themselves up to top ranks. Therefore, we decided against including kudo-ranks in our trustability function.

*Possible weakness of karma ranking.*

One must consider that developers may not use the same user names for all their commits through various repository systems. In such a case Ohloh can not auotmatically collect all the developers commits into one account; the developer would have to register and do this manually. Furthermore we blacklist commit bots. Finally the karma value could be tampered with deliberately if a user was to do a huge number of (small) commits to few highly ranked projects.

## 6. CONCLUSION

In this paper we have presented an approach to improve the *trustability* of search results. Trustability of search results is important, so that developers can quickly assess search results from external code bases before integrating them into their local code base.

We have proposed $T_p$ as a trustability metric for software projects. We have also presented *JBender*, a proof-of-concept prototype code search engine that implements the trustability metric which allows developers to quickly assess the trustability of search results from a code search engine. We have discussed the choice of our trustability metric and presented preliminary results from an ongoing evaluation.

The current trustability metric is defined per project. We would like to combine it with code ownership data from project history, so that we can assess the trustability of single classes (or even methods) based on developers karma.

Currently we are building up our metadata and code bases for *JBender*; upon reaching a sufficient level we plan do a user study to evaluate the effect of metadata on result trustability. We would also like to compare the proposed trustability metric with other trustability measurements, *e.g.*, corporate backing of projects. It might also be promising to combine the proposed trustability metric, which is currently based on human factors only, with technical trustability assessments such as *e.g.*, test coverage.

*Acknowledgments.*

We gratefully acknowledge the financial support of the Swiss National Science Foundation for the project "Bringing Models Closer to Code" (SNF Project No. 200020-121594, Oct. 2008 – Sept. 2010).